\documentstyle[twoside]{article}

\catcode`\@=11
\long\def\@makefntext#1{
\protect\noindent \hbox to 3.2pt {\hskip-.9pt  
$^{{\eightrm\@thefnmark}}$\hfil}#1\hfill}		

\def\@makefnmark{\hbox to 0pt{$^{\@thefnmark}$\hss}}	
	
\def\ps@myheadings{\let\@mkboth\@gobbletwo
\def\@oddhead{\hbox{}
\rightmark\hfil\eightrm\thepage}   
\def\@oddfoot{}\def\@evenhead{\eightrm\thepage\hfil
\leftmark\hbox{}}\def\@evenfoot{}
\def\sectionmark##1{}\def\subsectionmark##1{}}

\oddsidemargin=\evensidemargin
\newcounter{sectionc}\newcounter{subsectionc}\newcounter{subsubsectionc}
\renewcommand{\section}[1] {\vspace{12pt}\addtocounter{sectionc}{1} 
\setcounter{subsectionc}{0}\setcounter{subsubsectionc}{0}\noindent 
	{\tenbf\thesectionc. #1}\par\vspace{5pt}}
\renewcommand{\subsection}[1] {\vspace{12pt}\addtocounter{subsectionc}{1} 
	\setcounter{subsubsectionc}{0}\noindent 
	{\bf\thesectionc.\thesubsectionc. {\kern1pt \bfit #1}}\par\vspace{5pt}}
\renewcommand{\subsubsection}[1] {\vspace{12pt}\addtocounter{subsubsectionc}{1}
	\noindent{\tenrm\thesectionc.\thesubsectionc.\thesubsubsectionc.
	{\kern1pt \tenit #1}}\par\vspace{5pt}}

\newcommand{\textlineskip}{\baselineskip=13pt}
\newcommand{\smalllineskip}{\baselineskip=10pt}

\def\eightcirc{
\begin{picture}(0,0)
\put(4.4,1.8){\circle{6.5}}
\end{picture}}
\def\eightcopyright{\eightcirc\kern2.7pt\hbox{\eightrm c}} 

\newcommand{\copyrightheading}[1]
	{\vspace*{-2.5cm}\smalllineskip{\flushleft
        {\footnotesize 
        Comptes Rendus Acad. Sci. (Paris) 206 (1938) 1780-1782
        }\\
	 }}

\newcounter{itemlistc}
\newcounter{romanlistc}
\newcounter{alphlistc}
\newcounter{arabiclistc}

\def\@citex[#1]#2{\if@filesw\immediate\write\@auxout
	{\string\citation{#2}}\fi
\def\@citea{}\@cite{\@for\@citeb:=#2\do
	{\@citea\def\@citea{,}\@ifundefined
	{b@\@citeb}{{\bf ?}\@warning
	{Citation `\@citeb' on page \thepage \space undefined}}
	{\csname b@\@citeb\endcsname}}}{#1}}

\newif\if@cghi
\def\cite{\@cghitrue\@ifnextchar [{\@tempswatrue
	\@citex}{\@tempswafalse\@citex[]}}
\def\citelow{\@cghifalse\@ifnextchar [{\@tempswatrue
	\@citex}{\@tempswafalse\@citex[]}}
\def\@cite#1#2{{$\null^{#1}$\if@tempswa\typeout
	{IJCGA warning: optional citation argument 
	ignored: `#2'} \fi}}

\def\@refcitex[#1]#2{\if@filesw\immediate\write\@auxout
	{\string\citation{#2}}\fi
\def\@citea{}\@refcite{\@for\@citeb:=#2\do
	{\@citea\def\@citea{, }\@ifundefined
	{b@\@citeb}{{\bf ?}\@warning
	{Citation `\@citeb' on page \thepage \space undefined}}
	\hbox{\csname b@\@citeb\endcsname}}}{#1}}

\def\@refcite#1#2{{#1\if@tempswa\typeout
        {IJCGA warning: optional citation argument
	ignored: `#2'} \fi}}

\def\refcite{\@ifnextchar[{\@tempswatrue
	\@refcitex}{\@tempswafalse\@refcitex[]}}

\def\pmb#1{\setbox0=\hbox{#1}
	\kern-.025em\copy0\kern-\wd0
	\kern.05em\copy0\kern-\wd0
	\kern-.025em\raise.0433em\box0}

\def\fnt#1#2{\footnotetext{\kern-.3em
	{$^{\mbox{\scriptsize #1}}$}{#2}}}

\def\runninghead#1#2{\pagestyle{myheadings}
\markboth{{\protect\footnotesize\it{\quad #1}}\hfill}
{\hfill{\protect\footnotesize\it{#2\quad}}}}
\font\tenrm=cmr10
\font\tenit=cmti10 
\font\tenbf=cmbx10
\font\bfit=cmbxti10 at 10pt
\font\ninerm=cmr9

\font\eightrm=cmr8

\def\qed{\hbox{${\vcenter{\vbox{			
   \hrule height 0.4pt\hbox{\vrule width 0.4pt height 6pt
   \kern5pt\vrule width 0.4pt}\hrule height 0.4pt}}}$}}

\begin{document}

\pagestyle{empty}


\pagestyle{empty}

\begin{center}
{\it Los Alamos Electronic Archives: physics/9909061}
\end{center}

\bigskip

\begin{center}
{\bf MAKING OLD SEMINAL RESULTS WORLD-WIDE AVAILABLE !}
\end{center}

$\;$\\
$\;$\\
$\;$\\
$\;$\\
$\;$\\
$\;$\\

\begin{center}

{\bf FORWARD}

\end{center}

\bigskip

\noindent
Pioneering 1938 {\em Comptes Rendus} Paris Note of J. Delsarte on the
intertwining approach is archieved here for the Internet users.
One can find Delsarte's transformation operators (isomorphisms of
transmutations) for second-order partial differential equations briefly
presented in a general manner.
Only in the 1950s detailed studies of this approach followed that showed
its relevance in Physics.

\bigskip

\noindent
I would like to mention that a concept of {\em transference} has been
introduced by J.L. Burchenall and T.W. Chaundy in
Proc. London Soc. Ser. 2, {\bf 21},
420-440 (1923), but unfortunately I could not see this paper up to now.
According to M. Adler and J. Moser, {\em transferences} are Crum
transformations, which in turn are a simple form of intertwiners.

\bigskip

\noindent
For the benefit of the active authors and other interested people,
I offer the original French text of Delsarte's Note, together with
my personal English, Romanian and Spanish translations.

\bigskip
\bigskip

\hfill ${\cal H}$ ${\cal C}$ ${\cal R}$

\bigskip

\hfill 9. 29. 1999


\newpage

\runninghead{Jean Delsarte
$\ldots$} {Jean Delsarte
$\ldots$}


\normalsize\textlineskip
\pagestyle{empty}

\copyrightheading{}			

\vspace*{0.88truein}

\centerline{ANALYSE MATH\'EMATIQUE.}
\bigskip
\centerline{\bf SUR CERTAINES TRANSFORMATIONS FONCTIONELLES RELATIVES}
\centerline{\bf AUX \'EQUATIONS LIN\'EAIRES AUX D\'ERIV\'EES PARTIELLES DU
SECOND ORDRE}
\vspace*{0.035truein}
\vspace*{0.37truein}
\centerline{\footnotesize Note de {\it M}. J. Delsarte, pr\'esent\'ee par
{\it M}. Henri Villat; S\'eance du 13 Juin 1938}
\vspace*{0.015truein}
\centerline{[\footnotesize{\it en LaTex par} {\it M}. H.C. Rosu
(Septembre 1999)]}
\baselineskip=10pt
\vspace*{10pt}
\vspace*{0.225truein}

\vspace*{0.21truein}


\textlineskip                  
\vspace*{12pt}                 

\vspace*{1pt}\textlineskip	
\vspace*{-0.5pt}
\noindent


\noindent




Soit $R$ un nombre fixe; $A(r)$, $B(r)$, $C(r)$ seront trois fonctions
d\'efinies et continues pour $r\in (R,+\infty)$, la premi\`ere \'etant
essentiellement positive. Soient d'autre part $a(y)$, $b(y)$, $c(y)$
trois fonctions d\'efinies et continues pour $y\in (y_0;y_1))$.
Consid\'erons les \'equations
\begin{eqnarray*}
(1)&
\frac{\partial ^2f}{\partial x^2}= a(y)
\frac{\partial ^2f}{\partial y^2}+b(y)
\frac{\partial f}{\partial y} +c(y)f~,\\
(2)&
A(r)\frac{\partial ^2 F}{\partial r^2}+B(r)\frac{\partial F}{\partial r}
+C(r)F=a(y)
\frac{\partial ^2F}{\partial y^2}+b(y)
\frac{\partial F}{\partial y} +c(y)F~,\\
(3)&
A(r)\frac{\partial ^2 \Phi}{\partial r^2}+B(r)\frac{\partial \Phi}
{\partial r}+C(r)\Phi= \frac{\partial ^2 \Phi}{\partial t^2};
\end{eqnarray*}
dont on envisage respectivement les int\'egrales $f(x,y)$, $F(r,y)$,
$\Phi(r,t)$ d\'efinies et continues dans les domaines
\begin{eqnarray*}
D_1&:&\qquad \qquad x\in(-\infty, +\infty);\qquad \qquad y\in(y_0;y_1);\\
D_2&:&\qquad \qquad r\in(R, +\infty);\qquad \qquad y\in(y_0;y_1);\\
D_3&:&\qquad \qquad r\in(R, +\infty);\qquad \qquad t\in(-\infty;+\infty);~.
\end{eqnarray*}
Introduisons maintenant les quatre op\'erateurs lin\'eaires suivants
$$
f(r)={\cal A}_{r}[\alpha(t)];\qquad \qquad g(r)={\cal B}_{r}[\beta (\tau)]
$$
$$
\dot{\alpha}(t)=A_{t}[f(\rho)];\qquad \qquad  \dot{\beta}(t)=B_{t}[g(\rho)]~.
$$
Le premier donne la valeur $f(r)=\Phi(r,0)$ pour $t=0$, de l'int\'egrale
$\Phi(r,t)$ de (3), d\'efinie dans $D_3$ et satisfaisant aux conditions
$$
\Phi(R,t)=0,\qquad\qquad \left(\frac{\partial \Phi}{\partial r}
\right)_{r=R}=\alpha(t),\qquad\qquad t\in(-\infty, +\infty)~.
$$
Le second donne la valeur $g(r)=\Psi(r,0)$, pour $t=0$, de l'int\'egrale
$\Psi(r,t)$ de (3), d\'efinie dans $D_3$ et satisfaisant aux conditions
$$
\Psi(R,t)=\beta(t),\qquad\qquad
\left(\frac{\partial \Psi}{\partial r}\right)_{r=R}=0,
\qquad\qquad t\in(-\infty, +\infty)~.
$$
Le troisi\`eme donne la valeur $\alpha(t)=(\partial\Phi/\partial r)_{r=R}$,
de la d\'eriv\'ee par rapport \'a $r$, pour $r=R$, de l'int\'egrale
$\Phi(r,t)$ de (3), d\'efinie dans $D_3$ et satisfaisant
aux conditions
\[\left\{\begin{array}{ll}
\Phi(r,0)=f(r),&\\
& r\in(R,+\infty), \quad \Phi(R,t)=0, \quad t\in (-\infty,+\infty)~.\\
\left(\frac{\partial \Phi}{\partial t}\right)_{t=0}=0,
\end{array}
\right.
\]

Cette valeur est une fonction paire de $t$.

Le quatri\`eme donne la valeur $\dot{\beta}(t)=\Psi(R,t)$, pour $r=R$, de
l'int\'egrale $\Psi(r,t)$ de (3), d\'efinie dans $D_3$ et satisfaisant aux
conditions
\[\left\{\begin{array}{ll}
\Psi(r,0)=g(r),&\\
& r\in(R,+\infty), \quad \left(\frac{\partial \Psi}{\partial r}\right)_{r=R}=0,
\quad t\in (-\infty,+\infty)~.\\
\left(\frac{\partial \Psi}{\partial t}\right)_{t=0}=0,
\end{array}
\right.
\]

Cette valeur est une fonction paire de $t$; on notera que
$$
f(r)={\cal A}_{r}[\dot{\alpha}(t)]; \qquad \qquad g(r)={\cal B}_{r}
[\dot{\beta}(\tau)]~.
$$

\bigskip

Ceci \'etant, on peut \'enoncer les th\'eor\`emes suivants:

{\bf I}.
Si $f(x,y)$ et $g(x,y)$ sont des solutions de (1), d\'efinies et continues
dans $D_1$, les transformations
$$
F(r,y)={\cal A}_{r}[f(\xi,y)];\qquad \qquad G(r,y)={\cal B}_{r}[g(\xi,y)]
$$
leur font correspondre deux solutions $F(r,y)$ et $G(r,y)$ de l'\'equation
(2), d\'efinies et continues dans $D_2$.

\bigskip

{\bf II}.
Si $F(r,y)$ et $G(r,y)$ sont des solutions de l'\'equation (2), d\'efinies
et continues dans $D_2$, les transformations
$$
f(x,y)= A_{x}[F(\rho,y)];\qquad \qquad g(x,y)=B_{x}[G(\rho,y)]
$$
leur font correspondre deux fonctions paires de x, $f(x,y)$ et $g(x,y)$,
d\'efinies et continues dans $D_1$, et solution de l'\'equation (1).

\bigskip

{\em Exemple}. - (1) est l'\'equation des potentiels plans

$$
\frac{\partial ^2 f}{\partial x^2}+\frac{\partial ^2f}{\partial y^2}=0~,
$$
(2) est l'\'equation des potentiels r\'evolutifs
$$
\frac{\partial ^2 F}{\partial r^2}+\frac{1}{r}\frac{\partial F}{\partial r}
+\frac{\partial ^2F}{\partial y^2}=0~.
$$
Si $R$ est nul, on se trouve dans un cas limite, car alors les fonctions
$A(r)$, $B(r)$, $C(r)$ sont seulement d\'efinies et continues dans
$(0,+\infty)$; les op\'erateurs ${\cal A}$ et $A$ n'ont plus de sens;
on a \footnote{J'ai signal\'e la transformation correspondante
({\em Comptes rendus, 205, 1937, p. 645}).}
$$
{\cal B}_{r}[\beta(\tau)]=\frac{1}{\pi}\int _{-\pi/2}^{\pi/2}
\beta(r\sin \theta)d\theta~,
$$
$$
B_{t}[g(\rho)]=\frac{d}{dt}\Big[t\int _{0}^{\pi/2}g(t\sin \theta)\sin \theta
d\theta\Big]~.
$$
Si $R$ est positif, les quatre op\'erateurs ${\cal A}$, ${\cal B}$, $A$,
$B$ s'obtiennent ais\'ement sous forme finie par les proc\'ed\'es
classiques de la th\'eorie des \'equations hyperboliques; ils sont assez
compliqu\'es et font intervenir des int\'egrales dont les noyaux sont des
fonctions hyperg\'eom\'etriques.

\newpage



\centerline{Comptes Rendus Acad. Sci. Paris 206 (1938) 1780-1782}
\bigskip
\centerline{MATHEMATICAL ANALYSIS.}
\bigskip
\centerline{\bf ON SOME FUNCTIONAL TRANSFORMATIONS RELATIVE}
\centerline{\bf TO LINEAR PARTIAL DIFFERENTIAL EQUATIONS OF SECOND ORDER}
\vspace*{0.035truein}
\vspace*{0.37truein}
\centerline{\footnotesize Note by {\it Mr}. J. Delsarte, presented by
{\it Mr}. Henri Villat; Meeting of 13 Juin 1938}
\vspace*{0.015truein}
\centerline{[\footnotesize{\it in LaTex by} {\it Mr}. H.C. Rosu
(September 1999)]}
\baselineskip=10pt
\vspace*{10pt}
\vspace*{0.225truein}

\vspace*{0.21truein}


\textlineskip                  
\vspace*{12pt}                 

\vspace*{1pt}\textlineskip	
\vspace*{-0.5pt}
\noindent


\noindent




Let $R$ be a fixed number; $A(r)$, $B(r)$, $C(r)$ will be three functions
defined and continuous for $r\in (R,+\infty)$, the first being essentially
positive. On the other hand, let $a(y)$, $b(y)$, $c(y)$ be
three functions defined and continuous for $y\in (y_0;y_1))$.
Consider the equations
\begin{eqnarray*}
(1)&
\frac{\partial ^2f}{\partial x^2}= a(y)
\frac{\partial ^2f}{\partial y^2}+b(y)
\frac{\partial f}{\partial y} +c(y)f~,\\
(2)&
A(r)\frac{\partial ^2 F}{\partial r^2}+B(r)\frac{\partial F}{\partial r}
+C(r)F=a(y)
\frac{\partial ^2F}{\partial y^2}+b(y)
\frac{\partial F}{\partial y} +c(y)F~,\\
(3)&
A(r)\frac{\partial ^2 \Phi}{\partial r^2}+B(r)\frac{\partial \Phi}
{\partial r}+C(r)\Phi= \frac{\partial ^2 \Phi}{\partial t^2};
\end{eqnarray*}
where we focus on the integrals $f(x,y)$, $F(r,y)$,
$\Phi(r,t)$, respectively, defined and continuous in the domains
\begin{eqnarray*}
D_1&:&\qquad \qquad x\in(-\infty, +\infty);\qquad \qquad y\in(y_0;y_1);\\
D_2&:&\qquad \qquad r\in(R, +\infty);\qquad \qquad y\in(y_0;y_1);\\
D_3&:&\qquad \qquad r\in(R, +\infty);\qquad \qquad t\in(-\infty;+\infty);~.
\end{eqnarray*}
Let us introduce now the four linear operators as follows
$$
f(r)={\cal A}_{r}[\alpha(t)];\qquad \qquad g(r)={\cal B}_{r}[\beta (\tau)]
$$
$$
\dot{\alpha}(t)=A_{t}[f(\rho)];\qquad \qquad  \dot{\beta}(t)=B_{t}[g(\rho)]~.
$$
The first gives the value
$f(r)=\Phi(r,0)$ for $t=0$, of the integral
$\Phi(r,t)$ of (3), defined in $D_3$ and satisfying the conditions
$$
\Phi(R,t)=0,\qquad\qquad \left(\frac{\partial \Phi}{\partial r}
\right)_{r=R}=\alpha(t),\qquad\qquad t\in(-\infty, +\infty)~.
$$
The second gives the value $g(r)=\Psi(r,0)$, for $t=0$, of the integral
$\Psi(r,t)$ of (3), defined in $D_3$ and satisfying the conditions
$$
\Psi(R,t)=\beta(t),\qquad\qquad
\left(\frac{\partial \Psi}{\partial r}\right)_{r=R}=0,
\qquad\qquad t\in(-\infty, +\infty)~.
$$
The third gives the value
$\alpha(t)=(\partial\Phi/\partial r)_{r=R}$,
of the derivative with respect to $r$, for $r=R$, of the integral
$\Phi(r,t)$ of (3), defined in $D_3$ and satisfying the conditions
\[\left\{\begin{array}{ll}
\Phi(r,0)=f(r),&\\
& r\in(R,+\infty), \quad \Phi(R,t)=0, \quad t\in (-\infty,+\infty)~.\\
\left(\frac{\partial \Phi}{\partial t}\right)_{t=0}=0,
\end{array}
\right.
\]

$\alpha(t)$ is an even function of $t$.

The forth gives the value $\dot{\beta}(t)=\Psi(R,t)$, for $r=R$, of the
integral $\Psi(r,t)$ of (3), defined in $D_3$ and satisfying the
conditions
\[\left\{\begin{array}{ll}
\Psi(r,0)=g(r),&\\
& r\in(R,+\infty), \quad \left(\frac{\partial \Psi}{\partial r}\right)_{r=R}=0,
\quad t\in (-\infty,+\infty)~.\\
\left(\frac{\partial \Psi}{\partial t}\right)_{t=0}=0,
\end{array}
\right.
\]

$\dot{\beta}(t)$ is an even function of $t$; one can note the following
$$
f(r)={\cal A}_{r}[\dot{\alpha}(t)]; \qquad \qquad g(r)={\cal B}_{r}
[\dot{\beta}(\tau)]~.
$$

\bigskip

Given all the above, we can formulate the following theorems:

{\bf I}.
If $f(x,y)$ and $g(x,y)$ are solutions of (1), defined and continuous
in $D_1$, the transformations
$$
F(r,y)={\cal A}_{r}[f(\xi,y)];\qquad \qquad G(r,y)={\cal B}_{r}[g(\xi,y)]
$$
achieve a correspondence with two solutions $F(r,y)$ and $G(r,y)$,
respectively,
of the equation (2), which are defined and continuous in $D_2$.

\bigskip

{\bf II}.
If $F(r,y)$ and $G(r,y)$ are two solutions of the equation (2), defined and
continuous in $D_2$, the transformations
$$
f(x,y)= A_{x}[F(\rho,y)];\qquad \qquad g(x,y)=B_{x}[G(\rho,y)]
$$
achieve a correspondence with two even functions of $x$,
$f(x,y)$ and $g(x,y)$, respectively,
defined and continuous in $D_1$, and solution of the equation (1).

\bigskip

{\em Exemple}. - (1) is the potential equation in the plane

$$
\frac{\partial ^2 f}{\partial x^2}+\frac{\partial ^2f}{\partial y^2}=0~,
$$
(2) is the potential equation of cylindrical plane symmetry
$$
\frac{\partial ^2 F}{\partial r^2}+\frac{1}{r}\frac{\partial F}{\partial r}
+\frac{\partial ^2F}{\partial y^2}=0~.
$$
$R=0$ is a limiting case, because the functions
$A(r)$, $B(r)$, $C(r)$ are only defined and continuous in
$(0,+\infty)$; the operators ${\cal A}$ and $A$ have no meaning;
one has \footnote{I have already given the corresponding transformation in
{\em Comptes rendus, 205, 1937, p. 645}.}
$$
{\cal B}_{r}[\beta(\tau)]=\frac{1}{\pi}\int _{-\pi/2}^{\pi/2}
\beta(r\sin \theta)d\theta~,
$$
$$
B_{t}[g(\rho)]=\frac{d}{dt}\Big[t\int _{0}^{\pi/2}g(t\sin \theta)\sin \theta
d\theta\Big]~.
$$
If $R$ is positive, the four operators ${\cal A}$, ${\cal B}$, $A$,
$B$ can be easily obtained in explicit form by means of the classical
procedures of the theory of hyperbolic equations; they are quite complicated
and involve integrals with hypergeometric kernels.

\newpage



\centerline{Comptes Rendus Acad. Sci. Paris 206 (1938) 1780-1782}
\bigskip
\centerline{ANALIZ\u A MATEMATIC\u A}
\bigskip
\centerline{\bf ASUPRA ANUMITOR TRANSFORM\v ARI FUNC\c TIONALE RELATIVE}
\centerline{\bf LA ECUA\c TIILE LINEARE CU DERIVATE PAR\c TIALE SECUNDE}
\vspace*{0.035truein}
\vspace*{0.37truein}
\centerline{\footnotesize Not\u a a {\it Dl}. J. Delsarte, prezentat\v a de
{\it Dl}. Henri Villat; \c Sedin\c ta din 13 Iunie 1938}
\vspace*{0.015truein}
\centerline{[\footnotesize{\it \^{\i}n LaTex de  Dl}.
H.C. Rosu (Septembrie 1999)]}
\baselineskip=10pt
\vspace*{10pt}
\vspace*{0.225truein}


\textlineskip                  
\vspace*{12pt}                 

\vspace*{1pt}\textlineskip	
\vspace*{-0.5pt}
\noindent


\noindent




Fie $R$ un num\u ar fix; $A(r)$, $B(r)$, $C(r)$ vor fi trei func\c tii
definite \c si continue pentru $r\in (R,+\infty)$, prima fiind esen\c tial
pozitiv\u a. Fie, pe de alt\u a parte, $a(y)$, $b(y)$, $c(y)$
trei func\c tii definite \c si continue pentru $y\in (y_0;y_1)$.
S\u a consider\u am ecua\c tiile
\begin{eqnarray*}
(1)&
\frac{\partial ^2f}{\partial x^2}= a(y)
\frac{\partial ^2f}{\partial y^2}+b(y)
\frac{\partial f}{\partial y} +c(y)f~,\\
(2)&
A(r)\frac{\partial ^2 F}{\partial r^2}+B(r)\frac{\partial F}{\partial r}
+C(r)F=a(y)
\frac{\partial ^2F}{\partial y^2}+b(y)
\frac{\partial F}{\partial y} +c(y)F~,\\
(3)&
A(r)\frac{\partial ^2 \Phi}{\partial r^2}+B(r)\frac{\partial \Phi}
{\partial r}+C(r)\Phi= \frac{\partial ^2 \Phi}{\partial t^2};
\end{eqnarray*}
unde se eviden\c tiaz\u a respectiv integralele $f(x,y)$, $F(r,y)$,
$\Phi(r,t)$ definite \c si continue \^{\i}n domeniile
\begin{eqnarray*}
D_1&:&\qquad \qquad x\in(-\infty, +\infty);\qquad \qquad y\in(y_0;y_1);\\
D_2&:&\qquad \qquad r\in(R, +\infty);\qquad \qquad y\in(y_0;y_1);\\
D_3&:&\qquad \qquad r\in(R, +\infty);\qquad \qquad t\in(-\infty;+\infty);~.
\end{eqnarray*}
S\u a introducem acum urm\u atorii patru operatori lineari
$$
f(r)={\cal A}_{r}[\alpha(t)];\qquad \qquad g(r)={\cal B}_{r}[\beta (\tau)]
$$
$$
\dot{\alpha}(t)=A_{t}[f(\rho)];\qquad \qquad  \dot{\beta}(t)=B_{t}[g(\rho)]~.
$$
Primul indic\u a valoarea lui $f(r)=\Phi(r,0)$ pentru $t=0$,
a integralei
$\Phi(r,t)$ \^{\i}n (3), definit\u a \^{\i}n $D_3$
care satisface condi\c tiile
$$
\Phi(R,t)=0,\qquad\qquad \left(\frac{\partial \Phi}{\partial r}
\right)_{r=R}=\alpha(t),\qquad\qquad t\in(-\infty, +\infty)~.
$$
Al doilea indic\u a valoarea lui $g(r)=\Psi(r,0)$, pentru $t=0$,
a integralei
$\Psi(r,t)$ din (3), definit\u a \^{\i}n $D_3$ \c si care satisface
condi\c tiile
$$
\Psi(R,t)=\beta(t),\qquad\qquad
\left(\frac{\partial \Psi}{\partial r}\right)_{r=R}=0,
\qquad\qquad t\in(-\infty, +\infty)~.
$$
Al treilea indic\u a valoarea $\alpha(t)=(\partial\Phi/\partial r)_{r=R}$,
a derivatei \^{\i}n raport cu $r$, pentru $r=R$, a integralei
$\Phi(r,t)$ \^{\i}n (3), definit\u a \^{\i}n $D_3$ \c si satisf\u ac\^{\i}nd
condi\c tiile
\[\left\{\begin{array}{ll}
\Phi(r,0)=f(r),&\\
& r\in(R,+\infty), \quad \Phi(R,t)=0, \quad t\in (-\infty,+\infty)~.\\
\left(\frac{\partial \Phi}{\partial t}\right)_{t=0}=0,
\end{array}
\right.
\]

$\alpha (t)$ este o func\c tie par\u a de $t$.

Al patrulea indic\u a valoarea lui $\dot{\beta}(t)=\Psi(R,t)$, pentru $r=R$,
a integralei $\Psi(r,t)$ din (3), definit\u a \^{\i}n $D_3$ \c si
satisf\u ac\^{\i}nd condi\c tiile
\[\left\{\begin{array}{ll}
\Psi(r,0)=g(r),&\\
& r\in(R,+\infty), \quad \left(\frac{\partial \Psi}{\partial r}\right)_{r=R}=0,
\quad t\in (-\infty,+\infty)~.\\
\left(\frac{\partial \Psi}{\partial t}\right)_{t=0}=0,
\end{array}
\right.
\]

$\dot{\beta}(t)$ este o func\c tie par\u a de $t$; de notat c\u a
$$
f(r)={\cal A}_{r}[\dot{\alpha}(t)]; \qquad \qquad g(r)={\cal B}_{r}
[\dot{\beta}(\tau)]~.
$$

\bigskip

Toate acestea stabilite, se pot enun\c ta urm\u atoarele teoreme:

{\bf I}.
Dac\u a $f(x,y)$ \c si $g(x,y)$ sunt solu\c tii ale lui (1),
definite \c si continue
\^{\i}n $D_1$, transform\u arile
$$
F(r,y)={\cal A}_{r}[f(\xi,y)];\qquad \qquad G(r,y)={\cal B}_{r}[g(\xi,y)]
$$
le pun \^{\i}n coresponden\c t\u a dou\u a solu\c tii $F(r,y)$ \c si $G(r,y)$
ale ecua\c tiei
(2), definite \c si continue \^{\i}n $D_2$.

\bigskip

{\bf II}.
Dac\u a $F(r,y)$ \c si $G(r,y)$ sunt solu\c tii ale ecua\c tiei (2), definite
\c si continue \^{\i}n $D_2$, transform\u arile
$$
f(x,y)= A_{x}[F(\rho,y)];\qquad \qquad g(x,y)=B_{x}[G(\rho,y)]
$$
le pun \^{\i}n coresponden\c t\u a
dou\u a func\c tii pare de $x$, $f(x,y)$ \c si $g(x,y)$,
definite si continue \^{\i}n $D_1$, \c si solu\c tii ale ecua\c tiei (1).

\bigskip

{\em Exemplu}. - (1) este ecua\c tia poten\c tialelor plane

$$
\frac{\partial ^2 f}{\partial x^2}+\frac{\partial ^2f}{\partial y^2}=0~,
$$
(2) este ecua\c tia poten\c tialelor revolutive (cilindrice de simetrie
azimutal\u a)
$$
\frac{\partial ^2 F}{\partial r^2}+\frac{1}{r}\frac{\partial F}{\partial r}
+\frac{\partial ^2F}{\partial y^2}=0~.
$$
Dac\u a
$R$ este nul, ne g\u asim \^{\i}ntr-un caz limit\u a, pentru c\u a atunci
func\c tiile
$A(r)$, $B(r)$, $C(r)$ sunt definite \c si continue numai \^{\i}n
$(0,+\infty)$; operatorii ${\cal A}$ \c si $A$ sunt lipsi\c ti de sens;
in acest caz \footnote{Am semnalat transformarea corespunz\u atoare \^{\i}n
{\em Comptes rendus, 205, 1937, p. 645}.}
$$
{\cal B}_{r}[\beta(\tau)]=\frac{1}{\pi}\int _{-\pi/2}^{\pi/2}
\beta(r\sin \theta)d\theta~,
$$
$$
B_{t}[g(\rho)]=\frac{d}{dt}\Big[t\int _{0}^{\pi/2}g(t\sin \theta)\sin \theta
d\theta\Big]~.
$$
Dac\u a $R$ este pozitiv, cei patru operatori ${\cal A}$, ${\cal B}$, $A$,
$B$ se ob\c tin u\c sor \^{\i}n form\u a explicit\u a folosind procedeele
clasice
ale teoriei ecua\c tiilor hiperbolice; forma lor final\u a este destul de
complicat\u a
\c si \^{\i}n ele apar integrale cu nuclee care sunt
func\c tii hipergeometrice.

\newpage



\centerline{Comptes Rendus Acad. Sci. Paris 206 (1938) 1780-1782}
\bigskip
\centerline{AN\'ALISIS MATEM\'ATICO}
\bigskip
\centerline{\bf SOBRE ALGUNAS TRANSFORMACIONES FUNCIONALES RELATIVAS}
\centerline{\bf A LAS ECUACIONES LINEALES CON DERIVADAS PARCIALES SEGUNDAS}
\vspace*{0.035truein}
\vspace*{0.37truein}
\centerline{\footnotesize Nota de {\it Sr}. J. Delsarte, presentada por
{\it Sr}. Henri Villat; Junta de 13 Junio 1938}
\vspace*{0.015truein}
\centerline{[\footnotesize{\it Traducci\'on y LaTex por}
{\it Sr}. H.C. Rosu (Septiembre de 1999)]}
\baselineskip=10pt
\vspace*{10pt}
\vspace*{0.225truein}

\vspace*{0.21truein}


\textlineskip                  
\vspace*{12pt}                 

\vspace*{1pt}\textlineskip	
\vspace*{-0.5pt}
\noindent


\noindent




Sea $R$ un numero fijo; $A(r)$, $B(r)$, $C(r)$ ser\'an tres funciones
definidas y continuas para $r\in (R,+\infty)$, la primera siendo
esencialmente positiva. Sean, por otro lado, $a(y)$, $b(y)$, $c(y)$
tres funciones definidas y continuas para $y\in (y_0;y_1))$.
Consideremos las ecuaciones
\begin{eqnarray*}
(1)&
\frac{\partial ^2f}{\partial x^2}= a(y)
\frac{\partial ^2f}{\partial y^2}+b(y)
\frac{\partial f}{\partial y} +c(y)f~,\\
(2)&
A(r)\frac{\partial ^2 F}{\partial r^2}+B(r)\frac{\partial F}{\partial r}
+C(r)F=a(y)
\frac{\partial ^2F}{\partial y^2}+b(y)
\frac{\partial F}{\partial y} +c(y)F~,\\
(3)&
A(r)\frac{\partial ^2 \Phi}{\partial r^2}+B(r)\frac{\partial \Phi}
{\partial r}+C(r)\Phi= \frac{\partial ^2 \Phi}{\partial t^2};
\end{eqnarray*}
donde nos enfocamos a las integrales $f(x,y)$, $F(r,y)$,
$\Phi(r,t)$ definidas y continuas en los dominios
\begin{eqnarray*}
D_1&:&\qquad \qquad x\in(-\infty, +\infty);\qquad \qquad y\in(y_0;y_1);\\
D_2&:&\qquad \qquad r\in(R, +\infty);\qquad \qquad y\in(y_0;y_1);\\
D_3&:&\qquad \qquad r\in(R, +\infty);\qquad \qquad t\in(-\infty;+\infty);~.
\end{eqnarray*}
Se introducen ahora los cuatro siguientes operadores lineales
$$
f(r)={\cal A}_{r}[\alpha(t)];\qquad \qquad g(r)={\cal B}_{r}[\beta (\tau)]
$$
$$
\dot{\alpha}(t)=A_{t}[f(\rho)];\qquad \qquad  \dot{\beta}(t)=B_{t}[g(\rho)]~.
$$
El primero da el valor $f(r)=\Phi(r,0)$ para $t=0$ de la integral
$\Phi(r,t)$ de (3), definida en $D_3$ y satisfaciendo las condiciones
$$
\Phi(R,t)=0,\qquad\qquad \left(\frac{\partial \Phi}{\partial r}
\right)_{r=R}=\alpha(t),\qquad\qquad t\in(-\infty, +\infty)~.
$$
El segundo da el valor $g(r)=\Psi(r,0)$, para $t=0$, de la integral
$\Psi(r,t)$ de (3), definida en $D_3$ y satisfaciendo las condiciones
$$
\Psi(R,t)=\beta(t),\qquad\qquad
\left(\frac{\partial \Psi}{\partial r}\right)_{r=R}=0,
\qquad\qquad t\in(-\infty, +\infty)~.
$$
El tercero da el valor $\alpha(t)=(\partial\Phi/\partial r)_{r=R}$,
de la derivada respecto a $r$, en $r=R$, de la integral
$\Phi(r,t)$ de (3), definida en $D_3$ y satisfaciendo las condiciones
\[\left\{\begin{array}{ll}
\Phi(r,0)=f(r),&\\
& r\in(R,+\infty), \quad \Phi(R,t)=0, \quad t\in (-\infty,+\infty)~.\\
\left(\frac{\partial \Phi}{\partial t}\right)_{t=0}=0,
\end{array}
\right.
\]

$\alpha (t)$ es una funcion par de $t$.

El cuarto da el valor $\dot{\beta}(t)=\Psi(R,t)$, en $r=R$, de
la integral $\Psi(r,t)$ de (3), definida en $D_3$ y satisfaciendo las
condiciones
\[\left\{\begin{array}{ll}
\Psi(r,0)=g(r),&\\
& r\in(R,+\infty), \quad \left(\frac{\partial \Psi}{\partial r}\right)_{r=R}=0,
\quad t\in (-\infty,+\infty)~.\\
\left(\frac{\partial \Psi}{\partial t}\right)_{t=0}=0,
\end{array}
\right.
\]

$\dot{\beta}(t)$ es una funcion par de $t$; se puede notar lo siguiente
$$
f(r)={\cal A}_{r}[\dot{\alpha}(t)]; \qquad \qquad g(r)={\cal B}_{r}
[\dot{\beta}(\tau)]~.
$$

\bigskip

Con todo esto, se pueden enunciar los siguientes teoremas:

{\bf I}.
Si $f(x,y)$ y $g(x,y)$ son dos soluciones de (1), definidas y continuas
en $D_1$, las transformaciones
$$
F(r,y)={\cal A}_{r}[f(\xi,y)];\qquad \qquad G(r,y)={\cal B}_{r}[g(\xi,y)]
$$
logran poner en correspondencia a dos soluciones $F(r,y)$ y $G(r,y)$ de
la ecuacion (2), definidas y continuas en $D_2$.

\bigskip

{\bf II}.
Si $F(r,y)$ y $G(r,y)$ son dos soluciones de la ecuaci\'on (2), definidas
y continuas en $D_2$, las transformaciones
$$
f(x,y)= A_{x}[F(\rho,y)];\qquad \qquad g(x,y)=B_{x}[G(\rho,y)]
$$
logran poner en correspondencia a dos funciones pares
de $x$, $f(x,y)$ y $g(x,y)$,
definidas y continuas en $D_1$, y soluciones de la ecuaci\'on (1).

\bigskip

{\em Ejemplo}. - (1) es la ecuaci\'on de los potenciales planos

$$
\frac{\partial ^2 f}{\partial x^2}+\frac{\partial ^2f}{\partial y^2}=0~,
$$
(2) es la ecuaci\'on de los potenciales cil\'{\i}ndricos de simetr\'{\i}a
azimutal (revolutivos)
$$
\frac{\partial ^2 F}{\partial r^2}+\frac{1}{r}\frac{\partial F}{\partial r}
+\frac{\partial ^2F}{\partial y^2}=0~.
$$
Si $R$ es nulo, nos encontramos en un caso l\'{\i}mite, porqu\'e las funciones
$A(r)$, $B(r)$, $C(r)$ son definidas y continuas solamente en
$(0,+\infty)$; los operadores ${\cal A}$ y $A$ ya no tienen sentido;
tenemos \footnote{He se\~nalado la transformaci\'on correspondiente en
{\em Comptes rendus, 205, 1937, p. 645}.}
$$
{\cal B}_{r}[\beta(\tau)]=\frac{1}{\pi}\int _{-\pi/2}^{\pi/2}
\beta(r\sin \theta)d\theta~,
$$
$$
B_{t}[g(\rho)]=\frac{d}{dt}\Big[t\int _{0}^{\pi/2}g(t\sin \theta)\sin \theta
d\theta\Big]~.
$$
Si $R$ es positivo, los cuatro operadores ${\cal A}$, ${\cal B}$, $A$,
$B$ se pueden obtener facilmente en forma expl\'{\i}cita por los procedimientos
cl\'asicos de la teor\'{\i}a de las ecuaciones hiperb\'olicas; son bastante
complicados por la presencia de integrales con kernel en forma de
funciones hipergeom\'etricas.

\end{document}